\documentclass[aps,prr,twocolumn,superscriptaddress]{revtex4-2}

\usepackage{amssymb}
\usepackage{amsmath}
\usepackage{upgreek}
\usepackage{bm}
\usepackage{graphicx}
\usepackage{dcolumn}
\usepackage{longtable}
\usepackage{color}
\usepackage{xcolor}
\usepackage{hyperref}
\hypersetup{colorlinks = {true},
    		citecolor  = {blue},
    		urlcolor   = {blue},
    		linkcolor  = {blue}}

\newcommand{\Tc}{\ensuremath{T_{\rm c}}}

\newcommand{\EF}{\ensuremath{E_{\rm F}}}

\newcommand{\nuz}{\ensuremath{\nu_{z}}}

\newcommand{\GIT}{Ge$_{1-x}$In$_{x}$Te}

\newcommand{\Sup}{Supplemental Materials (SM)}
\newcommand{\SI}{SM}

\begin{document}


\author{Y.~Ihara}
\email[corresponding author: ]{yihara@phys.sci.hokudai.ac.jp}
\author{M. Shimohashi}
\affiliation{Department of Physics, Faculty of Science, Hokkaido University, Sapporo 060-0810, Japan}
\author{M. Kriener}
\email[corresponding author: ]{markus.kriener@riken.jp}
\affiliation{RIKEN Center for Emergent Matter Science (CEMS), Wako 351-0198, Japan}
\date{\today}

\title{Probing Mixed Valence States by Nuclear Spin-Spin Relaxation Time Measurements}

\begin{abstract}
Several elements in the periodic table exhibit an interesting and often overlooked feature: They skip certain valence states which is discussed in the field of superconductivity to be in favor of fostering higher transition temperatures $T_c$. 
However, from the experimental point of view, it is often deemed difficult to probe changes in the valence state. 
Here we demonstrate that the latter are accessible by the spin-spin relaxation rate $1/T_2$ in nuclear magnetic resonance. 
As target material, we chose the solid solution \GIT, where valence-skipping In induces superconductivity and changes its valence state as a function of $x$. 
We observe a strong enhancement in $1/T_2(x)$ and, most importantly, find that $1/T_2$ and $T_c$ exhibit a strikingly similar $x$ dependence. 
These results underline the importance of valence physics for the evolution of superconductivity in \GIT. 
A model based on a Ruderman-Kittel-Kasuya-Yosida type of interaction among the In nuclei is proposed which fully accounts for the experimental results.
\end{abstract}

\maketitle

\section{Introduction}

The valence-skipping degree of freedom is observed in several elements, such as Bi, Sn, In, Tl among others.
As an example, in prototypical BaBiO$_3$ a Bi ion should take its $4+$ state which means that a single electron occupies the $6s$ shell. 
This situation is known to be energetically unfavorable. 
Therefore Bi is considered to appear as a mixture of Bi$^{3+}$ and Bi$^{5+}$ in BaBiO$_3$ and, hence, Bi$^{4+}$ is the \textit{skipped} valence state. 
It was Varma \cite{varma88a} who firstly pointed out that this feature may be responsible for the high-temperature superconductivity (transition temperature $\Tc \sim 30 $~K) observed in K-doped BaBiO$_3$ \cite{cava88a}. However, to date there is no unambiguously clear experimental evidence that valence skipping indeed supports superconductivity although quite some experimental and theoretical work on various materials has been done over the years \cite{taraphder91a,themlin92a,taraphder95a,kazakov97a,armstrong99a,tsendin99a,dzero05a,matsushita05a,hase08a,ren13a,strand14a,sleight15a,plumb16a,hase16a,hase17a,wakita17a,kataria23a}. 
Apparently, one obstacle is the difficulty to capture the changes in the valence state as a function of an external parameter such as doping.
Among the methods employed are photoemission spectroscopy (PES), x-ray emission spectroscopy, M\"o\ss bauer spectroscopy, or the spin-lattice relaxation rate $1/T_1$ in nuclear magnetic resonance (NMR). \cite{themlin92a,armstrong99a,mito14a,wakita17a,kriener20a,mkim21a,nakanishi24a}
Since the energies of the different involved valence states are usually close to each other, the measured spectra are often broad and difficult to split quantitatively complicating their interpretation. 
Against this background, we propose to measure the NMR spin-spin relaxation rate $1/T_2$ rather than $1/T_1$ to study valence-state physics in solid solutions which is rather sensitive to the valence-mixing induced disorder, i.e., the degree of mixture of the different valence states.
To demonstrate the strength of this technique, we chose chemically simple \GIT\ as target material which has been reported to superconduct and where In is the valence-skipping element \cite{kriener20a,kriener22a}: 
When introduced into GeTe with the formal valence state Ge$^{2+}$ (cf.\ Refs.~\cite{hase16a,sleight09a} for this nomenclature), In replaces Ge$^{2+}$ and, hence, should also appear in the same valence state.
However, In$^{2+}$ is energetically unstable which creates the anticipated competition of its monovalent and trivalent states.   

In Ref.~\cite{kriener20a}, it is argued that initially trivalent In enters GeTe. 
In starts to appear also in its monovalent state when traversing $x\sim 0.12$ and the majority of the dopants becomes monovalent at intermediate $x\sim 0.4$. 
The authors also propose a phenomenological model based on this valence-state change which successfully accounts for all experimentally observed features. 
A brief summary of the impact of In doping into the polar semiconductor GeTe can be found in Section~S1 of the accompanying \Sup\ \cite{Suppl}.

The motivation of the present work is to test and further elucidate the valence-skipping scenario proposed in the literature by a different and complementary approach. 
Here we present a comprehensive $^{115}$In-NMR study on polycrystalline samples \GIT\ with $0 < x \leq 1$. 
We find a step-like change of the nuclear quadrupole resonance (NQR) frequency \nuz\ when traversing $x\sim 0.4$ which coincides with a  shift of the peak position in PES data on this system reported in \cite{kriener20a}.
Concomitantly, the nuclear spin-spin relaxation rate $1/T_2$ is strongly enhanced which is explained as indicative of the anticipated change in the In valence state. 
Based on these data, we propose a Ruderman-Kittel-Kasuya-Yosida (RKKY)-type interaction model among the In nuclear spins through the conduction electrons at \EF\ which accounts for the observed changes in $1/T_2$. 
The most important result of this work is that $1/T_2$ and \Tc\ exhibit a strikingly  similar $x$ dependence underlining the sensitivity and significance of $1/T_2$ measurements in valence-skipping materials. 
Moreover, this observation implies a close relationship between these two quantities and suggests that in \GIT\ the valence degree of freedom may be indeed one driving force for the observed enhancement of the superconducting \Tc\ with $x$. 

\section{Methods}

\subsection{Sample growth and characterization}
The \GIT\ batches were synthesized by conventional melt growth of stoichiometric mixtures of GeTe and InTe in evacuated quartz glass tubes.
These were fired for 48\,--\,72~h at 950$^{\circ}$C and then quenched into water.
For the batch with $x=0.08$, x-ray diffraction (XRD) data confirmed the rhombohedral $\alpha$-GeTe structure (space group $R3m$).
For all other batches with $0.25\leq x \leq 1$, XRD data of the obtained material indicates traces of the ambient-condition tetragonal InTe structure whose volume fraction quickly increases with $x$.
To obtain superconducting cubic material, approximately 400~mg of each prereacted batch were treated in a second growth step in which a high-pressure synthesis method was employed (5~GPa, 600\,--\,1300$^{\circ}$C), cf.\ Ref.~\cite{kriener20a} for the details.
The unit-cell volume shown in Fig.~\ref{fig1}(b) was extracted from the XRD data taken on each final batch. 
We note that the lifetime of the metastable cubic structure is more than one year for samples containing Ge and In \cite{kriener20a} and several months for pristine InTe \cite{banus63a}. 
To ensure the superconductivity, the $T_c$ value for $x=1$ was confirmed before and after the NMR experiment.

On selected batches the chemical composition was checked by energy-dispersive x-ray analysis.
It was found that the actual In concentration fits to the nominal one with the difference between them slightly increasing for $x \rightarrow 1$, cf.\ the discussion in Ref.~\cite{kriener20a}.
In the present work, the nominal composition is used when referring to a sample. 

\subsection{Measurements}
The superconductivity of the batches with $0.25\leq x \leq 1$ was confirmed by temperature $T$ dependent magnetization $M(T)$ measurements in a Quantum Design magnetic property measurement system (MPMS3) equipped with a $^3$He refrigerator on samples cut into a well-defined geometry.
Each sample was cooled down in zero magnetic field to the base temperature and $M(T)$ subsequently measured upon increasing temperature in an applied magnetic field $B=1$~mT.
The linear part of the magnetization below the transition was extrapolated and its intercept with the temperature axis defined as the superconducting \Tc.
The obtained \Tc\ values fit well to the published data in Ref.~\cite{kriener20a}.

The field-sweep NMR spectra were obtained by recording the fast-Fourier-transformed  NMR intensity during magnetic field ($\mu_0 H$) sweeps.
The NMR frequency was fixed to 41.51~MHz which corresponds to the reference field $\mu_0 H_{\rm ref} \approx 4.4$~T \cite{commentRefField}.
For the spin-echo pulse sequence, we utilized a shaped pulse to uniformly irradiate the radio-frequency field within the specified bandwidth of 400 kHz. \cite{ihara21a}
The measurement procedure of the spin-spin relaxation rate $1/T_2$ is described in detail in Section~S4 in the accompanying \SI\ \cite{Suppl}.

\section{Results}
\label{results}
\begin{figure}[t]
\centering
\includegraphics[width=0.9\linewidth]{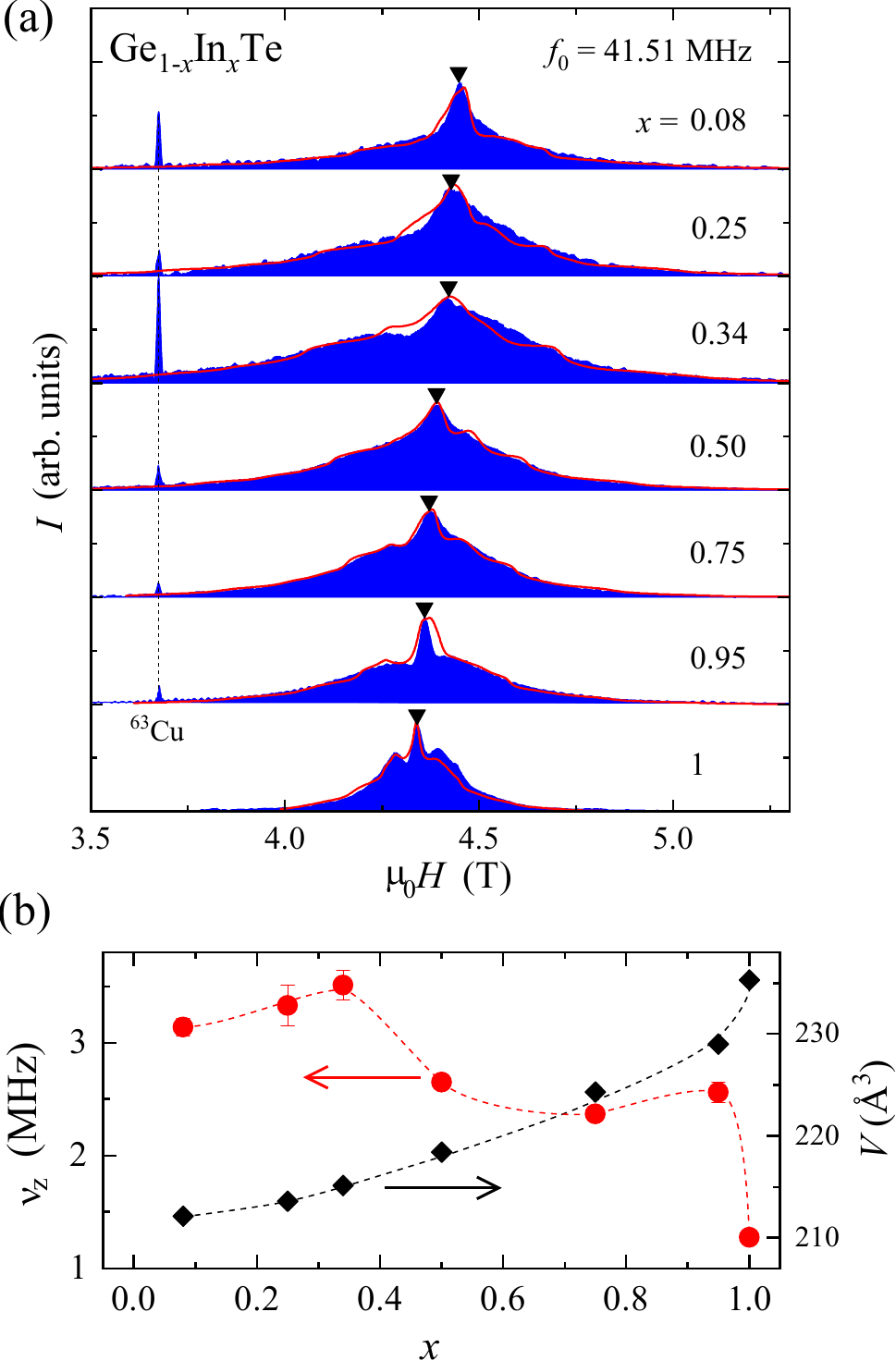}
\caption{
 (a) $^{115}$In-NMR spectra of \GIT\ on different samples taken at $T=4.2$~K ($0.08 \leq x \leq 0.95$) and at $T=1.6$ K ($x=1$). We note that the spectral shape does not exhibit any temperature dependence below 10 K.
 The spectra are vertically shifted for clarity. 
 The black triangles highlight the central peak positions and red curves are fits of the spectral shapes, see text.
The peaks at $\mu_0 H \sim 3.7$~T in the data for $x< 1$ are due to the Cu coil used in the NMR experiments. In the measurement of $x=1$ the narrower spectrum allowed to limit the field range to $ 3.8 \leq \mu_0 H \leq 5.0$~T explaining the absence of the Cu peak.
  (b) Extracted nuclear quadrupole resonance frequencies \nuz($x$) (red filled circles; left coordinate axis) and unit cell volume $V$ (black diamonds; right axis). The dashed lines are guides to the eyes.
}
\label{fig1}
\end{figure}

Figure~\ref{fig1}(a) summarizes $^{115}$In-NMR spectra of \GIT\.
All spectra consist of a central peak accompanied by broad tails on either side. 
The peak position evolves almost linearly with $x$, shifting toward slightly smaller fields as indicated by triangle symbols in Fig.~\ref{fig1}(a).

These NMR spectral shapes are understood as powder patterns for the $^{115}$In nuclear spin $I=9/2$ with allowing some asymmetry as follows: 
The nuclear spin Hamiltonian is given by the sum of the Zeeman and the electric quadrupolar interactions \cite{abragam83}:
\begin{equation*}
\mathcal{H}=-\gamma \hbar (1+K)\mu_0 \bm{H}_0\cdot \bm{I} + \frac{\nuz}{6}\left[ \left( 3I_z^2-\bm{I}^2 \right) + \frac{1}{2}\eta \left(I_{+}^2+I_{-}^2 \right) \right].
\end{equation*}
Here, $\gamma$ is the gyromagnetic ratio \cite{commentRefField}, $K$ is the Knight shift along the given orientation of the external field $\mu_0 \bm{H_0$}, $\eta$ represents the asymmetry of the electric field gradient (EFG) tensor around each In nuclear spin $\bm{I}$, and $\hbar$ the reduced Planck constant. 
The overall spectrum width is determined by the electric quadrupolar interaction, namely the NQR frequency $\nu_z$.  
This frequency is proportional to the largest EFG found along one of the three principal axes of the EFG tensor \cite{abragam83}.
The $^{115}$In-NMR spectra [Fig.~\ref{fig1}(a)] exhibit a slight broadening at intermediate doping, suggesting a systematic variation of $\nu_z$ as a function of $x$. 
To further analyze this, we fitted each powder pattern by optimizing the parameters $\nu_z$, $\eta$, and $K$. 
The results are shown as red curves in Fig.~\ref{fig1}(a) describing the data reasonably well; cf.\ also Section~S2 in \cite{Suppl}. 
We note that the finite $\eta = 0.3 \pm 0.1$ deduced from our fittings exhibits only a weak $x$ dependence.  

Figure~\ref{fig1}(b) (left axis) shows \nuz\ extracted for each $x$.  
Initially, \nuz\ increases with $x$.
The largest value of \nuz\ is found for the sample with $x=0.34$. 
Toward $x = 0.5$, \nuz\ is strongly suppressed and remains roughly constant above except for $x=1$, for which another abrupt decrease of \nuz\ is observed.
To understand the $x$ dependence of \nuz, one has to keep in mind that the EFG at the In sites is mainly determined by two different factors: 
(i) the crystal-lattice contributions from the ions surrounding an In site and (ii) the contributions due to interactions between charge carriers and nuclear In spins. 
The former, which should be zero when an In ion experiences perfect cubic symmetry, is caused by local distortions breaking the cubic symmetry around an In dopant when replacing a Ge ion. Hence, the In ions themselves create the EFG due to their different point charges as compared to Ge.
In the present case, this contribution is expected to lead to a monotonic suppression of \nuz\ with $x$ because the crystal lattice of \GIT\ expands monotonically as a function of $x$ up to $x = 1$, as shown in Fig.~\ref{fig1}(b) (right axis): 
The increasing distance between neighboring ions increases which in turn weakens the lattice EFGs. 
For pristine InTe with an ideal cubic structure, contributions from the lattice EFG are expected to vanish.  
Hence, the observed nonmonotonic $x$ dependence of $\nu_z$ and its finite value for $x=1$ imply that the interactions between the nuclear spins and the charge carriers produce a sizable contribution to the EFG, and thus, dominate $\nu_z(x)$.

Interestingly, the strong suppression of $\nuz(x)$ across $x\geq 0.34$ coincides with a change of the peak energy in In-3$d_{5/2}$ core-level PES data reported in Ref.~\cite{kriener20a}. 
Therein, this change was attributed to a crossover in the majority valence state of the In ions from trivalent to monovalent suggesting that the suppression in $\nuz(x)$ has the same origin.
However, the spectroscopic approach usually yields intrinsically broad spectra complicating the distinction of the involved mixed valence states.
\begin{figure}[t]
\centering
\includegraphics[width=0.9\linewidth]{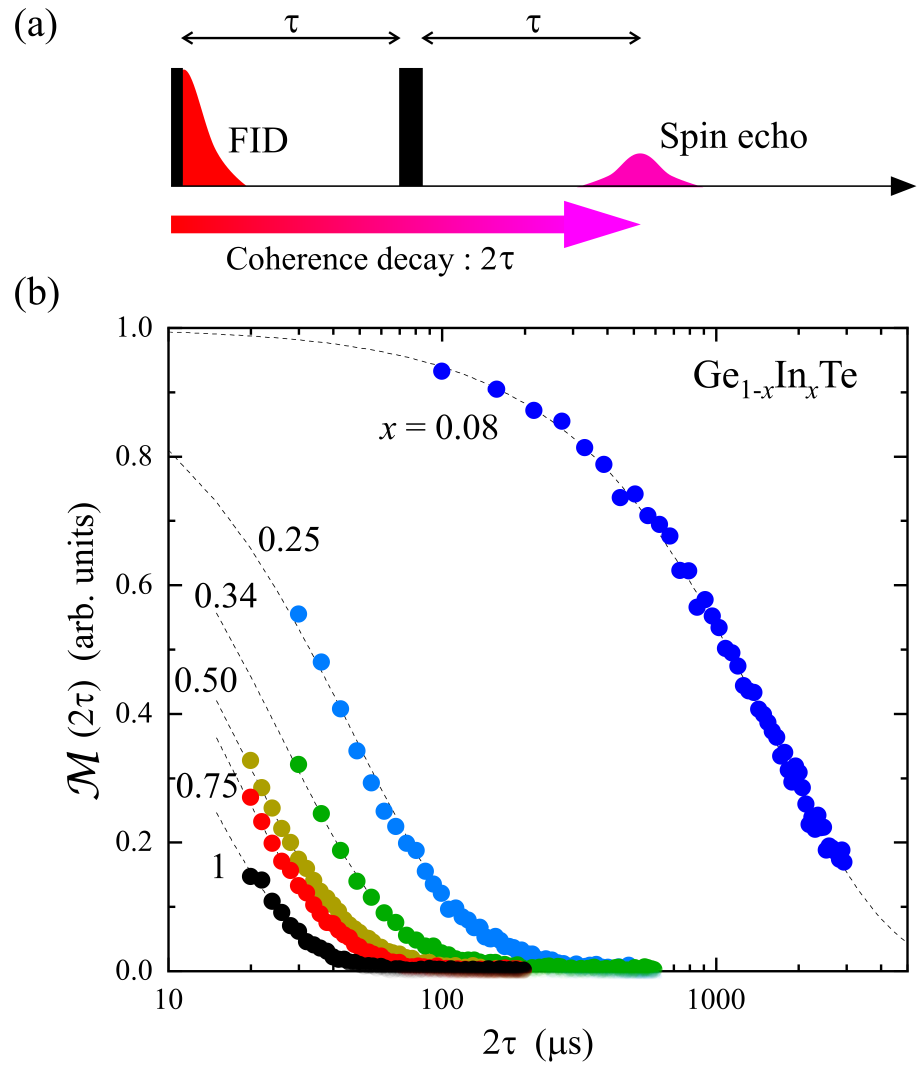}
\caption{
  (a) Sketch of spin-echo measurements and definition of the spin-echo pulse separation time $\tau$. 
 The free-induction decay (FID) signal observed after the first pulse disappears immediately and  
  the spin-echo signal appears at a time $\tau$ after the second pulse.  
(b) Spin-echo intensity $\mathcal{M}$ for \GIT\ with $0.08 \leq x \leq 1$ measured as a function of $\tau$.
Grey dotted lines are Lorentzian fits to the data.
The decay curves were measured at 4.2 K for $0.08\leq x \leq 0.75$ and at 1.6 K for $x=1$.
The decay curve for $x=0.95$ overlaps with that for $x=0.75$ and is not shown here for clarity. 
} 
\label{fig2}
\end{figure}
Therefore, we performed complementary nuclear spin-spin relaxation rate $1/T_2$ measurements to probe the effect of valence mixing, which are summarized in Fig.~\ref{fig2}. 
The relaxation time $T_2$ is the typical time scale on which nuclear spins loose their coherence in the interaction with their local environment.
Experimentally $1/T_2$ was determined by recording the spin-echo signal intensity $\mathcal{M}$ as a function of the separation time $\tau$ between two spin-echo pulses as sketched in Fig.~\ref{fig2}(a) \cite{commentT2}.  
Then the Lorentzian function 
$\mathcal{M}(\tau) = M_0\exp(-\tau/T_2)$ 
was fitted to each data set with the equilibrium magnetization $M_0$ of the undisturbed nuclear spin system, cf.\ \cite{abragam83} and Sections~S3 and S4 in the \SI\ \cite{Suppl}.
The results are shown as grey dashed lines in Fig.~\ref{fig2}(b) which reproduce the experimental data well. 
The extracted $x$ dependence of $1/T_2$ is replotted in Fig.~\ref{fig3}(a). 
The lowest-doped sample $x=0.08$ exhibits an extremely slow relaxation. 
Upon increasing $x$, $1/T_2$ is drastically enhanced for $x\leq 0.34$ and tends to saturate toward $x=0.95$. 
For $x=1$, a jump-like enhancement is observed.
Apparently, these features coincide with the suppression of $\nuz(x)$ across $x=0.34$, the approximately constant behavior up to $x=0.95$, and the subsequent strong suppression for $x=1$, cf.\ Fig.~\ref{fig1}(b).

Moreover, when comparing $1/T_2$ shown in Fig.~\ref{fig3}(a) with the superconducting \Tc\ displayed in Fig.~\ref{fig3}(b), a strikingly similar $x$ dependence of both quantities becomes apparent, suggesting a close correlation between them which will be discussed later.

\begin{figure}[t]
\centering
\includegraphics[width=0.9\linewidth]{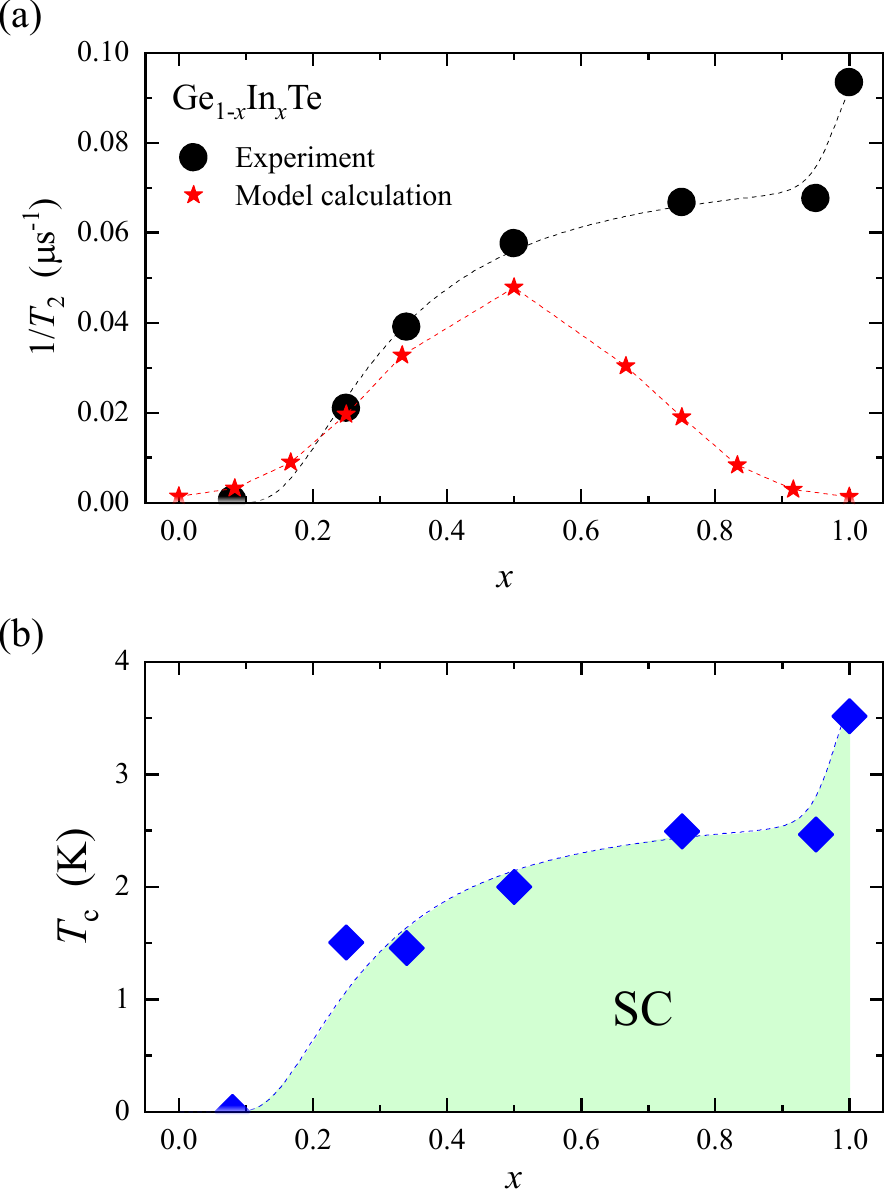}
\caption{
(a) In concentration $x$ dependence of $1/T_2$ deduced from the spin-echo decay curves (black filled circles) along with the results of model calculations (red stars), see text for details.
(b) Evolution of the superconducting transition temperature $T_c$. \cite{kriener20a}  
The superconducting phase (SC) emerges with $x$ and increases concomitantly with a change of the majority In valence state from In$^{3+}$ to In$^{1+}$, see text for details.
The dashed lines in both figures are guides to the eyes.}
\label{fig3}
\end{figure}

\section{Discussion}
\label{discussion}
\begin{figure*}[t]
\centering
\includegraphics[width=1\linewidth]{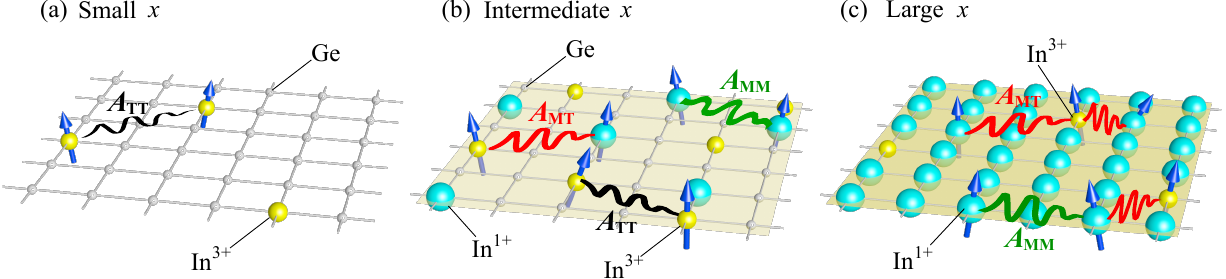}
\caption{Sketch of the situation in a (Ge,In) sublayer in \GIT\ for (a) small, (b) intermediate, and (c) large $x$. 
The grey, yellow, and blue balls represent Ge, In$^{3+}$, and In$^{1+}$, respectively. 
the background yellow shade depicts the concentration of the charge carriers, which increases as a function of $x$, cf.\ Section~S1 in the \SI\ \cite{Suppl}.
The In nuclear spins (blue arrows) are shown only on selected In sites for a better visibility.
Exchange interactions between different nuclear spins are indicated by black, green, and red bands and labeled $A_{\rm TT}$ (In$^{3+}$\,--\,In$^{3+}$; ``T'' = trivalent), $A_{\rm MM}$ (In$^{1+}$\,--\,In$^{1+}$; ``M'' = monovalent), and $A_{\rm MT}$ (In$^{1+}$\,--\,In$^{3+}$). 
At small $x$, only $A_{\rm TT}$ exchange is present because of the absence of In$^{1+}$.
This changes upon increasing $x$ leading to the appearance of the additional exchange interactions $A_{\rm MT}$ and $A_{\rm MM}$. 
Finally, at large $x$, the impact of the $A_{\rm TT}$ exchange vanishes because of the more and more increasing distances between the remaining In$^{3+}$ ions.
See text for details.}
\label{fig4}
\end{figure*}

To explain and contextualize the observed $x$ dependences of \nuz\ and $1/T_2$, one has to keep in mind that the decay rate $1/T_2$ is determined by the interaction between neighboring In nuclear spins. 
The initially very small value of $1/T_2$ and its strong increase toward intermediate $x$ cannot be explained by any simple dipole-dipole coupling, which is purely a function of the In\,--\,In distance and, hence, is expected to vary smoothly with $x$ \cite{abragam83}.
Therefore, another mechanism must be at work for which we propose an RKKY-type scenario based on the anticipated valence-state change with $x$ as illustrated in Fig.~\ref{fig4}.
Before discussing the details, we note that in an RKKY-type interaction, the conduction electrons at \EF\ mediate the scalar interaction between two nuclear spins $i$ and $j$ described by the Hamiltonian  $\mathcal{H}_{\rm RKKY}=A_{ij} \bm{I}_i  \cdot \bm{I}_j $.
The coupling constant $A_{ij}$ is determined by the hyperfine coupling strength between the conduction electrons and the In nuclear spins, the effective mass of the electrons, and the Fermi wave number of \GIT.
Hence, $A_{ij}$ is a function of the charge carrier concentration $n$ at \EF\ and, thus, the interaction between neighboring In nuclei strongly depends on $n$. 
If $\mathcal{H}_{\rm RKKY}$ and the nuclear spin operator $\bm{I}$ commute, the RKKY interaction does not destroy the quantum state of the nuclear spins.
This is the case if two interacting nuclear spins $\bm{I}_i$ and $\bm{I}_j$ are experiencing the same uniform local electronic state as it is the case when both nuclear spins belong to In ions with the same valence state.
This will not affect $1/T_2$.
However, the commutator $[\mathcal{H}_{\rm RKKY},\bm{I}]$ becomes finite when these interacting nuclear spins are in different electronic environments, i.e., experience different EFGs, which results in a shift in the resonance condition.
Such a situation is realized when the interacting nuclear spins belong to In dopants with different valence states and will significantly enhance $1/T_2$ \cite{commentRandom}.
Additional remarks concerning this RKKY-type interaction between nuclear spins can be found in Section~S5 in the \SI\ \cite{Suppl}.

Figure~\ref{fig4}(a) sketches the situation for small $x$:
Only a few In$^{3+}$ ions are embedded in the Ge-dominated background.
Since these are distributed homogeneously, they are too far away from each other and the carrier concentration \cite{kriener20a} is too small to allow for any sizeable RKKY-like interaction among their nuclear spins.
Hence, the respective coupling constant $A_{\rm TT}$ (``T'' = trivalent) is small.
When performing a spin-echo measurement at the In sites, each In nuclear spin resonates almost independently. 
As a result, $1/T_2 \rightarrow 0$ in agreement with our experimental observation for $x = 0.08$. 

This changes upon increasing $x$ as shown in Fig.~\ref{fig4}(b) which sketches the situation for intermediate $x$: 
The number of In nuclei and $n$ are larger \cite{kriener20a}, which enhances the coupling constant $A_{ij}$ and, hence, the RKKY interaction.
Moreover, now there are also In$^{1+}$ ions present resulting in two additional interaction paths with coupling constants $A_{\rm MM}$ and $A_{\rm MT}$ (``M'' = monovalent).
Since the nuclear spins of In$^{3+}$ ($\bm{I}_{\rm T}$) and In$^{1+}$ ($\bm{I}_{\rm M}$) experience a different electrical environment, as also reflected in the strong change of \nuz\ with $x$, the Hamiltonian 
$\mathcal{H}_{\rm RKKY}$ does not commute any more with $\bm{I}_{\rm T}$ and $\bm{I}_{\rm M}$. This leads to a faster decay in the spin-echo intensity which perfectly describes what is seen in Fig.~\ref{fig3}(a): $1/T_2$ is strongly enhanced toward intermediate $x$. 

Upon further increasing $x$, more and more cation lattice sites are occupied by In$^{1+}$ ions as illustrated in Fig.~\ref{fig4}(c). 
This provides a strong RKKY-type correlation among them mediated by a large number of conduction electrons at the Fermi level.
However, if all In ions were in their monovalent state at large $x$, $\mathcal{H}_{\rm RKKY}$ would commute again with the nuclear spin operators $\bm{I}_{\rm M}$ because all In nuclear spins would feel the same local electronic environment. Hence, in spite of the large prefactor $A_{\rm MM}$, $1/T_2$ should decrease toward large $x$ in contradiction to the experimental result.
To simulate this, we performed model calculations for $1/T_2$ based on the following scenario:
At small $x$, there are only In$^{3+}$ ions.
Upon increasing $x$, also In$^{1+}$ ions are present and the ratio of the populations of In$^{3+}$ and In$^{1+}$ changes from large to small as a function of $x$ ending for $x=1$ with In$^{1+}$ only, cf.\ also Section~S6 in the \SI\ \cite{Suppl}.
The results are shown in Fig.~\ref{fig3}(a) along with the experimental $1/T_2(x)$.
As expected, the simulation yields a dome-like $x$ dependence indicating that the decay of the nuclear-spin coherence is accelerated by the admixture of In$^{3+}$ and In$^{1+}$ peaking at intermediate doping.
Apparently, this does not agree with our experimental observation that $1/T_2$ remains large at large $x$.
This failure of the model calculations can only be explained if there are still sufficiently many In$^{3+}$ ions present for $x\rightarrow 1$, either statically or due to valence fluctuations, to allow for interactions with the In$^{1+}$ sites resulting in a large $1/T_2$. 
In Section~S7 in the \SI\ \cite{Suppl}, a rough estimation of the lower border for the required number of remaining In$^{3+}$ ions is given.

The final question to address is why $x=1$ is such an outlier: While both quantities $\nuz(x)$ and $1/T_2(x)$ are roughly constant for $x\rightarrow 1$, in pristine InTe the former is significantly suppressed and the latter strongly enhanced.  
This apparent impact on $\nuz(x)$ and $1/T_2(x)$ when only slightly changing the In content from $x=0.95$ to $x=1$ could indicate, that the crystal-lattice contributions to the EFG are different in \GIT\ for $x < 1$ and $x= 1$. 
While they are most likely negligible for $x < 1$, as argued above, this may change when divalent Ge is very diluted or completely absent.
The larger $1/T_2$ for $x=1$ also indicates that pure InTe is prone to possibly enhanced valence fluctuations as compared to $x<1$. 
Starting with InTe, Ge may introduce local lattice distortions which relax the frustration caused by valence-skipping In leading to a longer decay time for $x<1$ as compared to pristine InTe.
Remarkably, the observed $x$ dependence of the spin-echo relaxation rate $1/T_2$ [Fig.~\ref{fig3}(a)] almost perfectly resembles the $x$ dependence of the superconducting $T_c$ [Fig.~\ref{fig3}(b), \cite{kriener20a}] in the whole solid solution \GIT. This suggests an inherent coupling between both quantities and, hence, a close correlation between the valence-skipping feature of In and the evolution of the superconductivity in \GIT. Especially the additional strong enhancement of both when going from $x=0.95$ to $x=1$ possibly indicates that enhanced valence fluctuations are indeed supportive of the superconductivity as theoretically suggested \cite{varma88a} which is a tempting starting point for future studies on pristine InTe \cite{kriener22a}.

\section{Summary}
\label{summary}
This work reports a comprehensive $^{115}$In-NMR study on \GIT. 
We find a strong enhancement of the spin-spin relaxation rate $1/T_2$ when traversing $x = 0.34$ and its subsequent saturation.
This is explained as indicative of a change in the majority In valence state from trivalent to monovalent as a function of $x$. 
Concomitantly, we also observe a remarkably similar $x$ dependence of $1/T_2$ and the superconducting transition temperature $T_c$.
These findings point toward a scenario in which this valence-state change as a function of $x$ in \GIT\ and possibly enhanced valence fluctuations in InTe are beneficial for the superconductivity in this solid solution. 
Hence, this work provides strong indications that the valence-skipping feature of some elements can indeed be in favor of enhancing the superconducting pairing interaction as theoretically predicted.

The present work also demonstrates the high potential and power of approaching valence-state physics by means of spin-echo measurements, i.e., exploring electric properties of a solid with an essentially magnetic probe. 
This suggests to employ this technique also for other materials in which the valence degree of freedom may play a crucial role.

\section*{Acknowledgement}
This work was partly supported by JSPS KAKENHI (Grants No. 19H01832, No. 22H00104, No. 22H00263, and 22H04458). 
We thank M.~S.~Bahramy, K.~Ishida, D.~Maryenko, T.~Mito, H.~Mukuda, Y.~Nakai, J.~Ohara, Y.~Taguchi, and S.~Yonezawa for fruitful discussions.


\begin{thebibliography}{32}
\bibitem [{com({\natexlab{a}})}]{commentRefField}%
  \BibitemOpen
  \href@noop {} {}\bibinfo {howpublished} {{The gyromagnetic ratio of the
  $^{115}$In nuclear spin is $\gamma = 9.3295$ MHz/T, hence, $\mu_0 H_{\rm ref}
  \approx 4.4$~T.}}%
\bibitem [{com({\natexlab{b}})}]{commentT2}%
  \BibitemOpen
  \href@noop {} {}\bibinfo {howpublished} {{A more detailed description is
  given in Section~S4 in the \SI\ \cite{Suppl}.}} %
\bibitem [{com({\natexlab{c}})}]{commentRandom}%
  \BibitemOpen
  \href@noop {} {}\bibinfo {howpublished} {{This quantum mechanical result can
  be intuitively understood as follows: When all nuclear spins precess with the
  same Larmor frequency, the coherence of the resonance is preserved for an
  extended period of time. On the other hand, if two isospins precess with
  different frequencies, the dynamical interaction mediated by an RKKY
  mechanism will destroy the coherence.}} %
\end{thebibliography}
\end{document}


\title{Supplemental Materials for \\ ``Probing Mixed Valence States by \\ Nuclear Spin-Spin Relaxation Time Measurements''}

\author{Y.~Ihara}
\email[corresponding author: ]{yihara@phys.sci.hokudai.ac.jp}
\author{M. Shimohashi}
\affiliation{Department of Physics, Faculty of Science, Hokkaido University, Sapporo 060-0810, Japan}
\author{M. Kriener}
\email[corresponding author: ]{markus.kriener@riken.jp}
\affiliation{RIKEN Center for Emergent Matter Science (CEMS), Wako 351-0198, Japan}
\date{\today}

\maketitle
This \Sup\ provides additional details about \GIT\ and the measurement and analysis of the data as follows:

\begin{itemize}
\item Section~\ref{ppgit}: Evolution of structure and electronic parameters in \GIT
\item Section~\ref{psf}: Fitting of the powder spectra
\item Section~\ref{T2decay}: Spin-echo decay profile plotted on a linear scale
\item Section~\ref{T2m}: Measurement procedure of the spin-spin relaxation rate $1/T_2$
\item Section~\ref{rkky}: About the RKKY interaction between nuclear spins
\item Section~\ref{modcal}: Modeling of the mixture of In$^{3+}$ and In$^{1+}$ ions
\item Section~\ref{Tden}: Estimation of the In$^{3+}$ ion density for $x=1$
\end{itemize}

\clearpage

\begin{figure}[htbp]
\centering
\includegraphics[width=\textwidth]{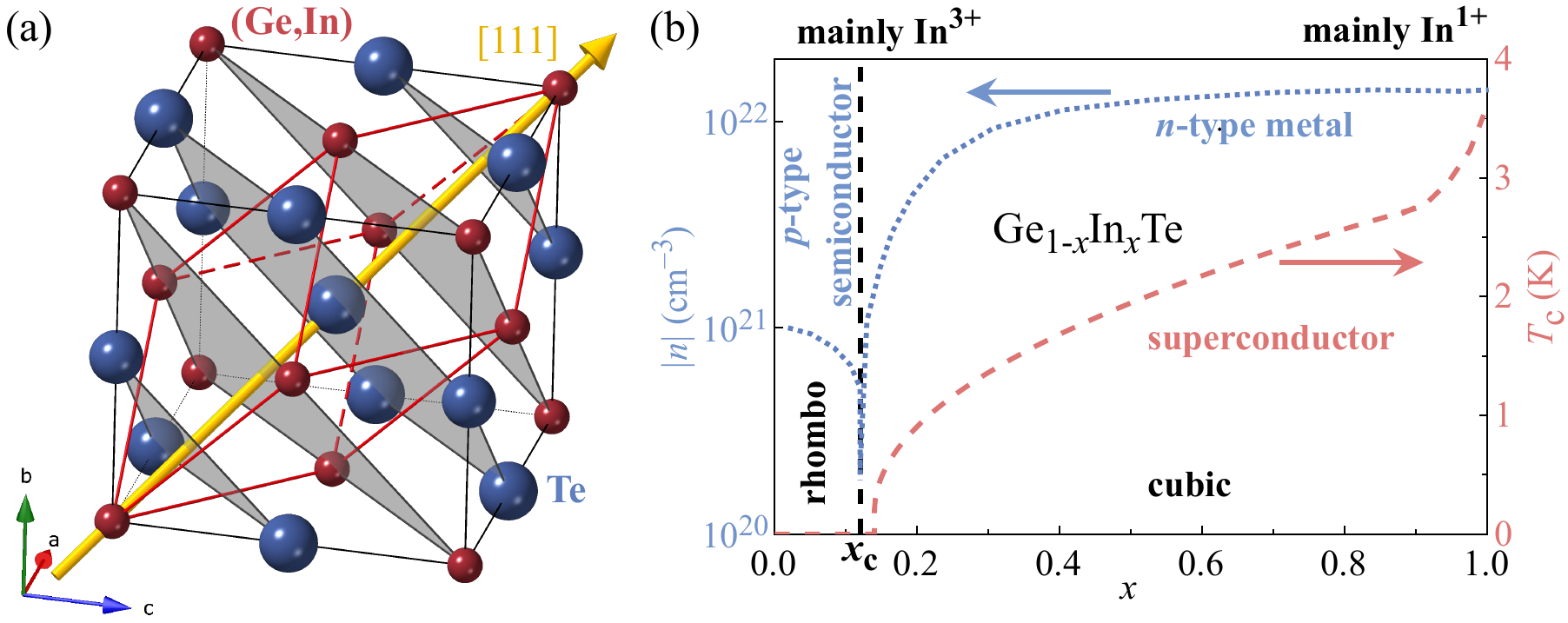}
\caption{
(a) Structural plot of and (b) sketch of the evolution of selected physical properties in the solid solution \GIT, see text for details.
}
\label{figS1}
\end{figure}

\subsection{Evolution of structure and electronic parameters in \GIT}
\label{ppgit}
Figure \ref{figS1}(a) provides a structural plot of \GIT~ and Fig. \ref{figS1}(b) summarizes the evolution of various physical parameters in the solid solution \GIT\ \cite{kriener20a}. 
In the latter, the $x$ dependence of the charge carrier concentration $|n|$ is plotted on the left vertical axis (blue), the right axis shows the superconducting transition temperature \Tc\ (red). The evolution of the crystal structure and the majority In valence state are also mentioned. The vertical dashed black line indicates the critical In concentration $x_{\rm c}=0.12$ across which various material properties change: 
For $x<x_{\rm c}$, the system is rhombohedrally distorted [space group $R3m$, no.\ 160; indicated as ``rhombo'' in Fig.~\ref{figS1}(b)]. 
The respective unit cell in its pseudocubic setting is highlighted in red in Fig.~\ref{figS1}(a). 
The distortion takes place along the cubic [111] direction indicated by the yellow arrow. 
At larger $x>x_{\rm c}$, the system crystalizes in a cubic structure ($Fm\bar{3}m$, no.\ 225; ``cubic'').
The corresponding unit cell is shown in black in Fig.~\ref{figS1}(a). 
As for the electric transport properties, pristine GeTe is a self-doped semiconductor with hole-type carriers (``$p$-type semiconductor''). Since In enters initially in its trivalent state (electron doping), the hole carriers are quickly compensated and the density of states is depleted making the system by far more insulating across $x=x_{\rm c}$. 
Upon further increasing $x$, the carrier type changes to electrons and the system becomes metallic (``$n$-type metal''). 
Moreover, a new superconducting phase emerges at slightly larger $x$ . 
In Ref.~\cite{kriener20a}, these features are explained by a crossover of the In valence state with $x$ from trivalent to monovalent as indicated on top of Fig.~\ref{figS1}(b). 
As discussed in the main text, this interpretation is supported by the present complementary and independent nuclear spin-spin relaxation time study on \GIT.

\begin{figure}[tbp]
\centering
\includegraphics[width=12cm]{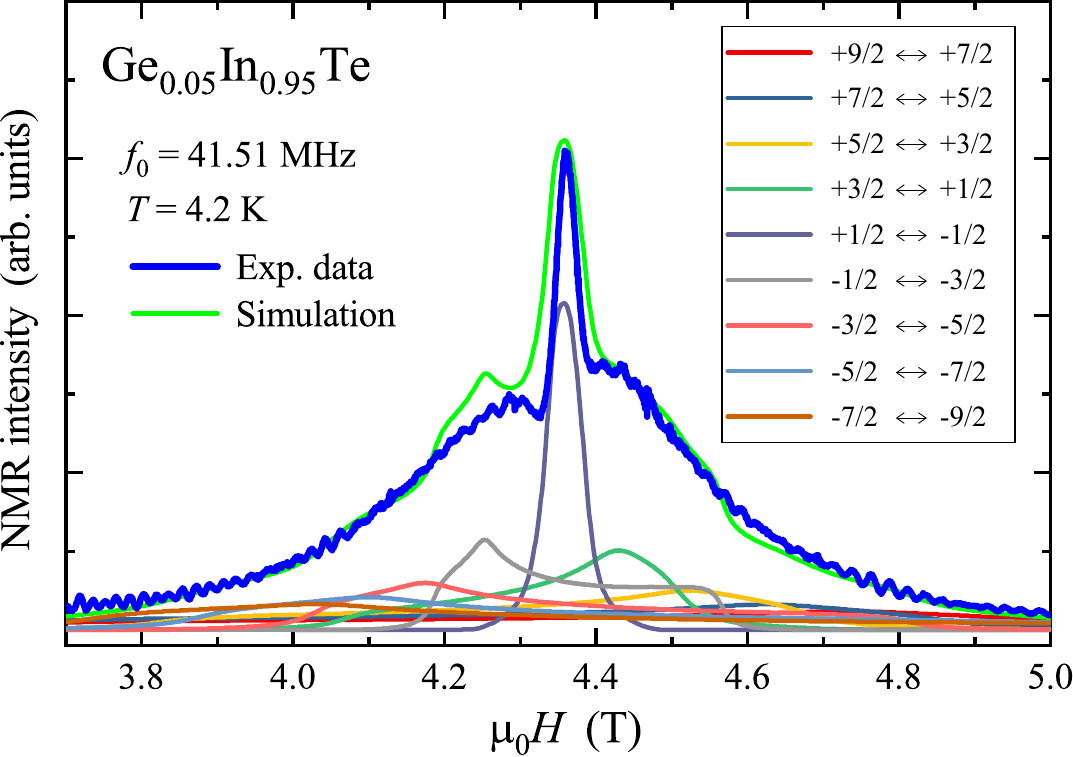}
\caption{
  Fitting of the quadrupolar-split powder pattern for the sample with $x=0.95$, see text for details.
}
\label{figS2}
\end{figure}

\subsection{Fitting of the powder spectra}
\label{psf}
The NMR spectrum of a powder sample is simulated by calculating the energy eigenvalues for nuclear spin states starting with the following nuclear spin Hamiltonian which is the sum of the Zeeman $\mathcal{H}_{\rm Z}$ and the electric quadrupolar $\mathcal{H}_{\rm Q}$ interactions\cite{abragam83}: 
\begin{align}
\mathcal{H} &= \mathcal{H}_{\rm Z} + \mathcal{H}_{\rm Q} \notag \\
&= - \gamma \hbar \left( 1+ K\right) \mu_0\bm{H}_0\cdot \bm{I} + \frac{\nu_z}{6} \left[ (3I_z^2-\bm{I}^2) +\frac{1}{2} \eta \left( I_+^2+I_-^2\right) \right]. 
\end{align}
Here, $K$, $\nu_z$, $\eta$, $\mu_0\bm{H}_0$, $\bm{I}$, and $\hbar$ are the Knight shift along the external field direction, the nuclear quadrupole resonance (NQR) frequency along the largest electric field gradient (EFG) direction ($z$ direction), 
the asymmetry parameter of the EFG tensor, the external field applied to the sample, the nuclear spin operator, and the reduced Planck constant, respectively. 
By using the second derivative of the electric potential $V$ along the $z$ direction $V_{zz} = \partial ^2 V/\partial z^2$, $\nu_z$ can be written as 
\begin{align}
 \nu_z = \frac{eQV_{zz}}{4I(2I-1)} , 
\end{align} 
where $e$, $Q$, and $I$ are the elementary charge, the nuclear quadrupole moment, and the nuclear spin, respectively. 
As the energy eigenvalues depend on the field orientation with respect to the principal axes of the EFG, the NMR peak positions also depend on the field orientation. 
The powder spectrum can be simulated by integrating the peak position over the magnetic field direction in a three-dimensional sphere. 

In the present study, the NMR measurements are performed at 41.51 MHz, which corresponds to the energy scale of the Zeeman interaction. Therefore, the quadrupole interaction, which is characterized by $\nu_z\simeq 3$ MHz, is treated as a perturbation to $\mathcal{H}_{\rm Z}$. 
In the simulations, the perturbation terms are included up to the second order. 

The experimentally obtained $^{115}$In-NMR spectrum of a sample consists of nine spectra originating from the transitions between the energy levels with the nuclear spin numbers $m_z \leftrightarrow m_z-1$ (with $m_z = +9/2, +7/2, \cdots -7/2$).
Figure~\ref{figS2} exemplarily shows a simulation for the sample with $x=0.95$: Summing up all nine single spectra yields the total spectrum (green line) which reasonably well reproduces the experimental data (blue): 
The sharp central peak ($m_z=+1/2 \leftrightarrow -1/2$) and the shoulders on either side are consistently reproduced by the simulation. 
In the present case ($x=0.95$), the simulation yields $\nu_z=2.32$ MHz and $\eta=0.3$. 
We note that the full width of the broad part of the spectrum is determined solely by $\nu_z$ irrespective of the small variation of $\eta = 0.3 \pm 0.1$, which only modifies the exact positions of the shoulders. 
The distribution of $\nu_z$ is taken into account by including a Lorentz-type broadening factor to the fitting procedure.
Broadening factors of 10 to 15\% with respect to \nuz were found to soften sharp kinks or shoulders otherwise present in the simulated spectra.
The doping dependence of $\nu_z$ is extracted from similar simulations performed for all other samples yielding $\nuz(x)$ plotted in Fig.~1(b) of the main text.

\begin{figure}[h]
\centering
\includegraphics[width=\textwidth]{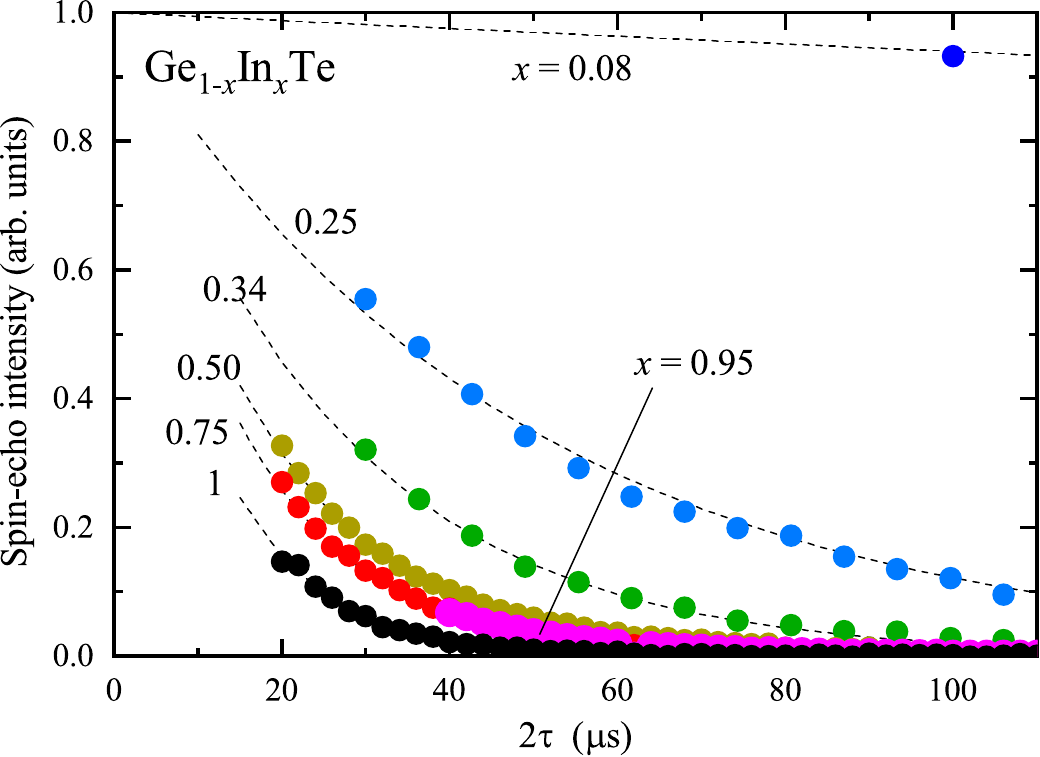}
\caption{
  Spin-echo decay profile plotted on a linear scale. The result for $x=0.95$, shown in magenta, overlaps with the decay curve for $x=0.75$ and is therefore omitted in Fig.~2(b) of the main text. 
}
\label{figS3}
\end{figure}

\subsection{Spin-echo decay profile plotted on a linear scale}
\label{T2decay}

In  Fig.~2(b) of the main text, the decay profiles of the spin-echo signals are plotted on a logarithmic scale to include the results for both the extremely long $T_2$ observed for $x=0.08$ and the rather short $T_2$ for $x=1$. 
Figure~\ref{figS3} presents the same data on a linear scale to better visualize the variation of the decay curves for short $T_2$ (large $x$).
Here, also the data for $x=0.95$ are shown. These are omitted in Fig.~2(b) of the main text for clarity because they overlap with the result for $x=0.75$.    
The identical recovery profiles for $x=0.75$ and $x=0.95$ and the apparently faster decay for $x=1$ confirm the saturating behavior of $1/T_2$ above $x\sim 0.5$ and point toward an abrupt increase in the case of pure InTe.

\begin{figure}[tbp]
\centering
\includegraphics[width=\textwidth]{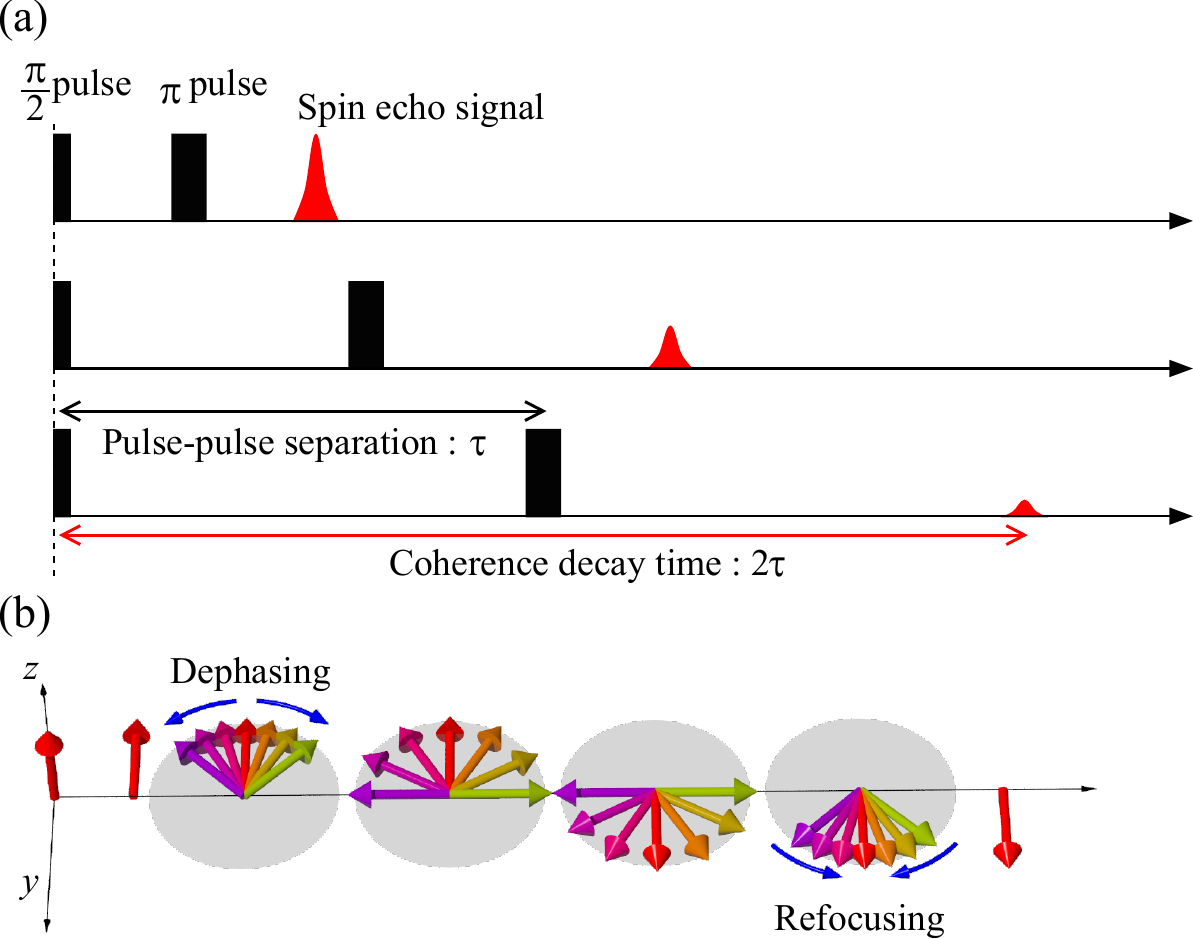}
\caption{
 (a) Schematic showing the pulse sequence typically employed in the spin-echo method to determine $T_2$.
 (b) The arrows represent classical nuclear moments throughout the spin-echo pulse sequence shown in (a).
}
\label{figS4}
\end{figure}

\subsection{Measurement procedure of the spin-spin relaxation rate \boldmath{$1/T_2$}}
\label{T2m}

The nuclear spin-spin relaxation time $T_2$ is the typical timescale for the resonating nuclear-spin ensemble to lose its coherence. 
The coherent precession motion is disrupted either by a static magnetic field distribution or by dynamical field fluctuations. 
With the spin-echo method, one can selectively measure the dynamical contribution. 

The radio frequency (rf) pulse sequence and the subsequent spin-echo signal are illustrated in Fig.~\ref{figS4}. 
In an external magnetic field $H_0$, the nuclear magnetization points along the field direction. 
When irradiating a $\pi/2$ pulse with the frequency $\omega_0=\gamma H_0$ ($\gamma = 9.3295$ MHz/T is the gyromagnetic ratio of the $^{115}$In nuclear spin) at the beginning, the nuclear magnetization rotates by 90 degrees. 
After the $\pi/2$ pulse, the nuclear spin ensemble dephases because of the static distribution of the magnetic field, and the in-plane component of the nuclear magnetization diminishes to zero within a short time. 
After the disappearance of the in-plane nuclear magnetization, a $\pi$ pulse is irradiated after the time $\tau$ has passed. This causes each nuclear spin to rotate by 180 degrees, i.e., their initial orientation is restored by following the same path they took before the $\pi$ pulse was applied. 
Finally, the dephased nuclear magnetization will be refocused at $2\tau$, yielding the so-called spin-echo signal. 
As the dephasing effect due to the static field distribution is corrected by this refocusing mechanism, 
the spin-echo intensity decreases only due to the change in the magnetic-field distribution before and after the $\pi$ pulse.

Experimentally, $T_2$ can be measured by recording the spin-echo intensity as a function of $2\tau$. 
The spin-echo intensity decay is described either by the Lorentzian function 
\begin{equation}
M_{\rm Lorentz}(2\tau )=M_0\exp \left( -2\tau/T_2 \right)
\end{equation}
in case of a weak interaction or by the Gaussian function 
 \begin{equation}
M_{\rm Gauss}(2\tau ) = M_0\exp \left( -(2\tau/T_2)^2 \right)
\end{equation}
 in case of a strong interaction. 
Practically, $T_2$ is extracted by fitting the measured decay data to one of these functions. 
In principle, the decay curve transforms from the Lorentzian to the Gaussian type when $T_2$ becomes shorter due to stronger interactions. 
In this study, however, the experimental decay curves of all samples are fitted by the Lorentzian function $M_{\rm Lorentz}(2\tau)$ 
because the fitting range is limited to $2\tau>20$ $\upmu$s, and, therefore, 
it is not possible to describe the decay curves for the whole $2\tau$ measurement interval in the case that $T_2$ becomes extremely short which is the case here for large $x$.
Thus, for consistency the same functional form was used for all samples to determine the variation of $T_2$ as a function of $x$.

\subsection{About the RKKY interaction between nuclear spins}
\label{rkky}

\subsubsection*{\textbf{The coupling constant \boldmath{$A_{ij}$}}}

In a metallic material, nuclear spin-spin interactions are indirectly mediated by the conduction electrons \cite{ruderman54a}. 
The electronic spin scattered by a nuclear spin $I_i$ at $R_i$ propagates to another nuclear spin $I_j$ at $R_j$ 
and modifies the spin state through a second-order perturbation effect.  
This mechanism is exactly the same as the indirect interaction between localized impurity spins embedded in metals \cite{boyce76a}.
In the case of an interaction between nuclear spins, 
carrier doping increases the RKKY interaction strength through the concomitant increase of the hyperfine coupling constant $H_{\rm hf}$ between the nuclear spins and the conduction electrons, 
which is proportional to the carrier density at the nuclear spin site $\rho(0) = |\psi_s|^2$.  
The Hamiltonian for the anticipated RKKY-type interaction takes the scalar form 
$\mathcal{H}_{\rm RKKY} = A_{ij}\bm{I}_i\cdot \bm{I}_j$.  
The coefficient $A_{ij}$ is determined by $H_{\rm hf}$, the effective mass of the electrons $m^\ast$, and the Fermi wave number \kF\ ($\hbar$ denotes the reduced Planck constant): 
\begin{align}
  A_{ij} &= \frac{m^\ast}{8\pi^3} \left( \frac{H_{\rm hf}}{\hbar}\right)^2 \frac{2\kF r_{ij}\cos (2\kF r_{ij})-\sin (2\kF r_{ij})}{r_{ij}^4}. 
\end{align} 
We note that $A_{ij}$ oscillates as a function of the relative position $r_{ij}=R_i-R_j$ and decays as $1/r_{ij}^3$.  
When crossing $x_{\rm c} = 0.12$, the RKKY interaction strengthens due to the carrier doping into a rigid band because of the increase of $H_{\rm hf}$. 
When the band structure has changed more significantly with $x$ due to the increasing number of In dopants, $A_{ij}$ is also more strongly modified due to the concomitant changes in $m^\ast$ and \kF.
To correctly estimate the doping dependence of $A_{ij}(x)$ in Ge$_x$In$_{1-x}$Te, the band structure for each $x$ is needed, which is rather difficult to obtain in a first-principles approach.
Therefore, for simplicity, in this study $A_{ij}$ is assumed to be constant for all $x$ and we rather focus qualitatively on the effect of mixing different nuclear spins, 
which is essential to understand the doping dependence of $1/T_2$ in a system where different valence states are simultaneously present.

\subsubsection*{\textbf{Impact of the RKKY interaction on \boldmath{$1/T_2$}}}
As mentioned above, the nuclear spin-spin interaction $\mathcal{H}_{\rm RKKY}$ in the anticipated RKKY mechanism takes a scalar form.
In this Section we describe the conditions required for $\mathcal{H}_{\rm RKKY}$ to disrupt the spin coherence and, thus, accelerate the decay rate $1/T_2$. 
As argued in the previous section, the quantitative discussion of the doping dependence of $A_{ij}$ is beyond the scope of this work because it requires to know the band structure for each $x$.  
Therefore, we focus here on a qualitative feature that originates from the basic principles of quantum mechanics.

The time evolution of the in-plane nuclear magnetization $m_y(t)$ depends on the density matrix $\rho(t)$ as $m_y(t) = {\rm Tr}\left[ \rho(t) I_y \right]$.
Here, $\rho(t)$ is written by the unitary matrix $U(t) = \exp(-i\mathcal{H}t/ \hbar)$ ($\mathcal{H}$ denotes the total Hamiltonian) as
\begin{align}
\rho(t) = U(t) \rho(0) U^{-1}(t). 
\end{align}
We define $\rho(0)$ as the density matrix right after the $\pi/2$ pulse which is proportional to $I_y$. 
A sketch showing the evolution of classical nuclear moments during a spin-echo pulse measurement is shown in Fig.~\ref{figS4}(b).
During the time interval $0<t<\tau$, $\rho(t)$ evolves according to 
\begin{align}
\mathcal{H} = -\gamma \hbar (H_0+\delta H) \cdot I_z+ A_{ij}\bm{I}_i\cdot \bm{I}_j,
\end{align}
where $\delta H$ denotes the static field distribution. 
The $\pi$ pulse at $t=\tau$ transforms $I_y \rightarrow -I_y$ and $I_z \rightarrow -I_z$, 
and the time evolution of $\rho(t)$ after $t=\tau$ continues according to the transformed Hamiltonian $\mathcal{H}'$. 

Now, if the commutator $[\mathcal{H}_{\rm RKKY}, \bm{I}_i]=0$, the time evolution due to the Zeeman interaction is canceled out which allows to calculate the density matrix at $t=2\tau$ as follows:
\begin{align}
\label{eq4}
\rho(2\tau) = \exp \left( -i \frac{\mathcal{H}_{\rm RKKY}}{\hbar} 2\tau \right) I_y \exp \left( i \frac{\mathcal{H}_{\rm RKKY}}{\hbar} 2\tau \right). 
\end{align}
The $\tau$ dependence of $\rho(2\tau)$ is further canceled when the RKKY interaction is isotropic, i.e., in this case there is no effect on the spin-echo decay. 
By contrast, if $[\mathcal{H}_{\rm RKKY}, \bm{I}_i] \neq 0$, the time evolution at $t<\tau$ will never be canceled after the $\pi$ pulse, and thus, 
the finite RKKY interaction accelerates the decay of the spin-echo intensity. 

The Hamiltonian $\mathcal{H}_{\rm RKKY}$ and $\bm{I}_i$ commute when $\bm{I}_i$ and $\bm{I}_j$ are in the same electric environment, i.e., both operators can be diagonalized simultaneously. 
By contrast, even if the quantum numbers and magnetic moments of the spins $\bm{I}_i$ and $\bm{I}_j$ are the same, they cannot be diagonalized together if the NMR frequencies for these spins are different. Thus, the commutator is finite. 
In the present study, as the NMR frequency for the $^{115}$In spins of the In$^{3+}$ and In$^{1+}$ ions are different 
because of the difference in their local electric environment,
the RKKY interaction accelerates the decay of the spin-echo intensity in the $x$ range where In ions with both valence states are simultaneously present. 

\subsection{Modeling of the mixture of In\boldmath{$^{3+}$} and In\boldmath{$^{1+}$} ions}
\label{modcal}
We assume a spin chain with 12 spins to study the effect of the RKKY interaction on the spin-echo decay. 
We also assume that for $x\rightarrow0$, all In ions are in their trivalent state and for $x\rightarrow 1$ all take their monovalent state.
The total Hamiltonian is written as 
\begin{align}
  \mathcal{H} = - \sum g \mu_i \bm{I}_i \cdot \bm{B} + \sum A_{ij} \mu_i \bm{I}_i \cdot \mu_j \bm{I}_j.
\end{align}
Here, site-dependent magnetic moments $\mu_i$ are introduced to simulate the mixing effect of In ions with different valence states. 
For small $x$, all In ions feature a uniform $\mu$ because there are only In$^{3+}$ ions present. Upon increasing $x>x_{\rm c}$, more and more In ions take their monovalent state and, hence, more and more spin sites are replaced with $\mu'$. Eventually, all In ions are In$^{1+}$ for $x\rightarrow 1$, i.e., again all In ions exhibit a uniform $\mu'$.
The total Hamiltonian is diagonalized and the density matrix given in Eq.~\ref{eq4} is numerically calculated using the respective wave function for each $x$.  
The resulting $2\tau$ dependence of $\rho(2\tau)$ is fitted to the Lorentzian function in order to extract theoretical $T_2$ values yielding the $x$ dependence of $1/T_2$ shown in Fig.~3(a) of the main text after normalizing it by an arbitrary factor to be readily compared with the experimental result. 
While our model calculations qualitatively reproduce the experimentally observed $x$ dependence of $1/T_2$ for $x\leq 0.5$, this is not the case for larger $x$. Hence, the assumption that all In ions are monovalent when approaching $x=1$ is wrong.

\subsection{Estimation of the In\boldmath{$^{3+}$} ion density for \boldmath{$x=1$}}
\label{Tden}
As argued in the last Section, to explain the observed behavior of $1/T_2$ at large $x$, a certain fraction of the In ions has to be in their trivalent state even for $x=1$. This gives rise to a finite RKKY interaction $A_{\rm MT}$ between the nuclear spins of In ions with different valence states to allow for a continued disturbance of the coherence of their nuclear spins. 
Here, we present a simple formula to estimate the density required to trigger an interaction among the majority of the In$^{1+}$ sites. 
Assuming that the RKKY interaction between an In$^{3+}$ ion and the surrounding In$^{1+}$ ions extends over a sphere with radius $l_0$ measured from the central In$^{3+}$ site, the number $N$ of In$^{1+}$ in this sphere is given by 
\begin{equation}
  N = \frac{4\pi}{3}l_0^3 \frac{1}{4 a^3} - 1 
\end{equation}
with the cubic lattice constant $a$ of \GIT. 
Measuring $l_0$ in multiples of the lattice constant $l_0=n\cdot a$, the density $p$ of the In$^{3+}$ ions required is 
\begin{equation}
  p = \frac{1}{\pi n^3/3-1}.
\end{equation}
Assuming an interaction width of $n=2.5$ lattice constants, we obtain that $p\approx 7$\% In$^{3+}$ ions are sufficient to guarantee a finite interaction $A_{\rm MT}$.


%